# Modelling collective motion based on the principle of agency

Katja Ried,[1]* Thomas Müller,[2]† Hans J. Briegel[1,2]‡

[1]*Institut für Theoretische Physik, Universität Innsbruck, Technikerstraße 21a, 6020 Innsbruck, Austria*
[2]*Department of Philosophy, University of Konstanz, 78457 Konstanz, Germany*

December 4, 2017

**Abstract**

Collective motion is an intriguing phenomenon, especially considering that it arises from a set of simple rules governing local interactions between individuals. In theoretical models, these rules are normally *assumed* to take a particular form, possibly constrained by heuristic arguments. We propose a new class of models, which describe the individuals as *agents*, capable of deciding for themselves how to act and learning from their experiences. The local interaction rules do not need to be postulated in this model, since they *emerge* from the learning process. We apply this ansatz to a concrete scenario involving marching locusts, in order to model the phenomenon of density-dependent alignment. We show that our learning agent-based model can account for a Fokker-Planck equation that describes the collective motion and, most notably, that the agents can learn the appropriate local interactions, requiring no strong previous assumptions on their form. These results suggest that learning agent-based models are a powerful tool for studying a broader class of problems involving collective motion and animal agency in general.

## 1 Introduction

Collective behaviour is a wide-spread phenomenon in biology: fish school to reduce drag, large herbivores gather in herds to avoid predation, bees distribute the task of locating a new nesting site, and even bacteria employ collective decision-making mechanisms in certain circumstances [1]. A case of particular interest are locusts, short-horned grasshoppers which form swarms that can consume hundreds of tons of food per day and threaten the livelihood – in the case of a plague – of up to a tenth of the world's population [2].

In an effort to understand the dynamics of locusts swarms and ultimately develop more efficient methods to control outbreaks, a number of experimental studies have been conducted. For instance, Buhl *et al.* observed marching locusts in a laboratory setting, using computer vision to track individuals and study the small-scale movement and interactions that ultimately give rise to swarming [3]. On the theoretical side, a range of models have been proposed, which can be roughly divided into two classes. On the one hand, coarse-grained models provide a concise description of the dynamics of the entire swarm in terms of global parameters such as the average alignment [4, 5]. This approach allows for an efficient description regardless of the size of the swarm and can account for striking features of the collective dynamics.

However, a better understanding of the origin of the swarm behaviour arguably requires a more fine-grained model, which explicitly describes interactions at the level of individuals. Such models typically describe the members of a swarm as self-propelled particles (SPPs): that is, particles capable of altering their trajectories in response to the motion of other nearby particles, usually governed by a set of simple interaction rules. The seminal SPP model was proposed by Vicsek *et al.*, who demonstrated that a population of particles which move with a fixed speed and tend to align themselves with their neighbours can give rise to several distinct regimes of collective dynamics [6]. Since then, many variations of SPP models have been studied, some of them explicitly designed to model marching locusts: for example, the pause-and-go model [7], which was inspired by the empirical observation that marching locusts frequently interrupt their movement, allows the particles to switch between moving with fixed speed $v_0$ and staying still. The escape-and-pursuit model [8] embodies the thesis that cannibalistic interactions play a crucial role in the emergence of collective motion by including two interaction terms, encoding the complementary tendencies to avoid potential attackers but to approach potential victims. A noteworthy extension of this model was explored by Guttal et al., who combined the escape-and-pursuit interaction with genetic algorithms to find the weights of the two interaction terms in a way that models evolutionary adaptation [9].

*Katja.Ried@uibk.ac.at
†Thomas.Mueller@uni-konstanz.de
‡Hans.Briegel@uibk.ac.at

Considering the existing individual-based models of collective motion in locusts, one may note that they generally make strong assumptions regarding the form of the interaction between individuals: in Vicsek-style models, particles generally update their speeds to gradually approach the average speed of their neighbours, while in zonal avoidance-alignment-attraction models [10, 11, 12], for example, they switch among a small set of fixed interaction rules (e.g. embodying either attraction or avoidance) when a neighbour crosses certain thresholds in terms of distance or relative speed. Crucially, the form of these interactions – while justified with heuristic arguments – is essentially proposed *ad hoc*.[1]

Moreover, in order to account for the existence of different regimes of collective motion, some models postulate that there are several different interaction rules (or at least different values of relevant parameters) which are toggled in different circumstances, e.g. depending on the swarm density [4] or on the distance between individuals [11]. Indeed, it is known that the responses of individual locusts to surrounding conspecifics can change, most notably in the transition from solitarious to gregarious phase. However, the object of greatest interest (at least for the present work) are locusts which are already in the gregarious phase, in which case it would be more appropriate to uncover a single, *stable* rule governing individual behaviour that explains circumstance-dependent changes in their behaviour, rather than postulating a set of different rules that apply under different circumstances.

In the present work, we propose a more general framework for modelling the collective motion of locusts which addresses both of these issues. The defining feature of our model is that it treats members of the swarm as *learning agents*. While the term 'agent' (as in 'agent-based') has been used before, as a synonym of 'individual', for example in the context of generic SPP-type models, we take it to carry a strictly stronger meaning: an agent has an explicit *perception* of its surroundings, to which it responds according to a single, *situation-independent* set of rules, and it is able to *adapt* its responses based on its personal history of interactions and feedback. This notion of an agent is grounded in philosophical considerations on the nature of agency, but agents of this type have also proved to be of great practical value for the purpose of learning in the context of artificial intelligence and robotics. Inspired by the usefulness of the concept of agents in these fields, the present work explores what insights learning agent-based modelling can provide in the study of collective motion, in particular regarding the phenomenon of density-dependent alignment in marching locusts.

The remainder of the paper is organised as follows: Section 2 introduces the framework of learning agents and a formal description of the scenario which we will study. In section 3, we begin by considering agents with fixed behavioural tendencies, which allows us to show that a a single (fixed) individual-level interaction rule can account for different regimes of collective motion. In section 4, we show that, in a limiting case, these results can be described by a Fokker-Planck equation, which allows for a straightforward comparison with the predictions of other models. Section 5 focuses on the effects of learning, whereby agents can develop suitable interaction rules autonomously instead of having to be provided with them by a programmer. Finally, section 6 discusses the implications of our findings and provides an outlook at possible directions for further work.

---

[1] Once an individual-level interaction rule has been proposed, one can test whether the effects that it produces at the level of collective motion are in agreement with experimental observations. (For a description and illustration of this procedure, see e.g. [5].) However, the underlying individual-level rules are generally not checked directly against empirical observations.

# 2 Fundamentals

## 2.1 The framework of learning agents

Given the central role that learning agents play in our model, we begin with some fundamental considerations of what sets such agents apart and how they can be modelled formally. The first essential element of an agent-based model is a description of how the agent interacts with its environment. On the one hand, the agent obtains information about the environment in the form of a *percept*s, which we denote $s$. In response to a percept, the agent chooses an *action a*, which in turn changes the state of the environment. For the purpose of book-keeping, the interaction can be divided into turns, which gives rise to an alternating sequence of percepts and actions. The idea of modelling an agent's perception explicitly may prove particularly useful in the study of collective behaviour in animals, since it allows one to explore which stimuli individuals must be able to perceive in order to account for the observed behaviour.

The second part of an agent-based model describes the agent's internal processes, by which it deliberates on the received percept, chooses an action, and eventually updates the mechanisms of deliberation. Research in both neuroscience and artificial intelligence has generated a variety of proposals for how these internal processes could work, some of which can account for remarkable feats of learning and intelligent behaviour. However, the more sophisticated of these models generally require relatively powerful universal computers to run, and therefore cannot be considered accurate models of how biological entities – whose neural architecture is, at the most fundamental level, relatively simple – can exhibit such behaviours. We favour instead a simpler model of the agent's internal processes, one that does not presume general-purpose computational capabilities, but which is instead supported by the natural dynamics of the physical system that embodies these processes. In the case of locusts in particular, this restriction to relatively basic internal processes seems realistic. Moreover, even if the actual agent is a more complex organism, simpler models quite generally yield more easily interpretable results.

A particularly important feature of the agent's internal processes is its capacity to retain a memory of past interactions with the environment: this enables the agent to gradually develop its own, individual behavioural tendencies, and therefore sets it apart from mere pre-programmed automata. In order for us to assess whether an agent has this type of memory and learning ability, it helps if the agent has a clear goal to achieve. For example, one can easily conclude that an animal is learning if it modifies its behaviour in such a way as to navigate a maze – and ultimately reach a reward – more quickly. In the case of artificial agents, the reward is reduced to an abstract variable, but it nevertheless plays an essential role in the learning process: upon receiving a reward, the agent updates its deliberation mechanism, reinforcing those responses that eventually lead to a reward.[2] Over time, these changes improve the agent's chances of earning rewards, and thus the agent learns.

A concrete proposal for an agent of this type was introduced by Briegel and De las Cuevas in [13]. The mathematical and conceptual framework developed therein is termed Projective Simulation (PS), after the defining feature of the agents: they can effectively project themselves into conceivable futures and simulate likely consequences of their actions before realizing them. While such high-level capabilities are not necessary for the present work, a simpler variant of the PS agent does promise to be an interesting model for individuals that can perceive their neighbours, respond to their presence according to certain – adaptable – interaction[3] rules and ultimately exhibit collective dynamics.

The interaction of a PS agent with its environment follows the scheme introduced above. Its memory structure, which supports the internal deliberation process, is based on snippets of previous experiences and actions, termed *clips*. Transitions between clips are interpreted as the agent recalling or simulating sequences of events that have been rewarded, for instance "percept: predator" → "action: flight". Formally, one can represent the clips as vertices and the possible transitions as edges of a directed graph, which is termed the *clip network* or episodic and compositional memory (ECM, see [13]). Simple agents possess only two types of clips, representing remembered percepts and actions, which are arranged in a two-layer network. In general, more sophisticated agents may also possess intermediate clips that represent neither pure percepts nor actions, but this added complexity is not required for the present task. The deliberation process of a PS agent is realized as a random walk over the clip network, with an excitation hopping from clip to clip along the edges of the graph. The walk is governed by the *hopping matrix h*, which stores information about past interactions in the form of *h-values* $h_{ij} \geq 1$ associated with each edge $i \to j$. In the standard PS framework, the probability of the excitation hopping from a clip $i$ to a clip $j$ is determined from the $h$-values by simple normalization,

$$P(j|i) = \frac{h_{ij}}{\sum_k h_{ik}}. \tag{1}$$

---

[2] Notice that the agent does not need to retain an explicit memory of the sequence of all percepts, actions and rewards that it has encountered. Instead the agent can build implicit memories, e.g. by strengthening connections between percepts and actions that led to rewards, which only store the information that is relevant for the task of earning more rewards in the future.

[3] In the context of collective behaviour, one needs to distinguish between interaction rules and learning rules. By interaction rules, in the present work, we mean those rules that describe how an individual interacts with its neighbours. These rules generally change over the course of a learning process. The learning rules, on the other hand, which describe how the agent processes rewards, are embodied and therefore in principle fixed.

Once an action clip is reached, the agent realizes the corresponding action. If a reward is received, in the simplest case, its numerical value $R$ is added to the entries in the $h$-matrix that correspond to the transitions used in the random walk, thereby making these transitions more likely in the future. (The value $R = 0$ therefore corresponds to no reward being received.) More generally, an agent may need to adapt to a variety of learning scenarios that present different challenges, such as reward schemes that change over time or delayed rewards. By refining the above rules for deliberation and updating, it is possible to endow the agent with the ability to handle several more general classes of problems without actually tailoring it to particular tasks. One refinement that is relevant for the present work addresses the possibility of an environment that changes with time by enabling the agent to *forget* previously acquired information. This can be implemented by damping the $h$-matrix by a factor of $1 - \gamma$: if $\gamma = 0$, the agent does not forget at all; if $\gamma = 1$, it forgets immediately and is therefore incapable of learning over time. Combining these mechanisms, the update rule reads

$$h_{ij}^{(t+1)} - 1 = (1 - \gamma)\left(h_{ij}^{(t)} - 1\right) + \begin{cases} R^{(t)} & \text{for used transitions} \\ 0 & \text{for unused transitions.} \end{cases}, \tag{2}$$

where $R^{(t)}$ denotes the reward received at turn $t$.

As far as its learning dynamics is concerned, projective simulation belongs to the class of reinforcement learning models. Indeed, one could also use standard reinforcement learning algorithms, such as Q-learning or SARSA [14], to study the learning process that allows us to derive the interaction rules. One of the benefits of using the PS model is that it offers an explicit and straightforward account of the agent's deliberation dynamics. This should become particularly important for the modelling of agents with higher cognitive capacities.

## 2.2 The phenomenon of density-dependent alignment

The present work focuses on one particular feature of the collective dynamics of marching locusts, namely their density-dependent alignment [3]: at low densities, the motion is completely disordered, whereas at high densities, individual movement is strongly aligned in a single direction, which remains constant for long periods of time. At intermediate densities, the group is also strongly aligned, but switches direction roughly periodically.

In order to simplify the mathematical description of the phenomenon, it is convenient to constrain the locusts to move (approximately) in one dimension, for instance in a narrow circular arena, which minimizes transversal movement. In this case, the alignment can be described by a single parameter $z$, which measures the group average of the direction of motion: it ranges from +1, if all individuals move in the clockwise direction, to -1, indicating perfect counterclockwise alignment. Plotting the time-dependence $z(t)$ reveals a striking visual signature of three regimes of collective motion, as one can see e.g. in the experimental data reported in [3]. This visual signature can also be seen in the results of our own simulations, in Fig. 1.

Following existing practice, the world in which the agents move is taken to be one-dimensional and circular, with the ends identified so that the agents can cross in both directions. The world is divided into $W$ discrete spatial blocks. The agents move at a fixed speed of one block per time-step and can pass through each other (that is, we do not restrict the number of agents occupying the same position). A second fundamental parameter in our model is the sensory range $r$, which gives the distance up to which a given agent can perceive others. Together with the size $W$ of the world, this defines the number $B = W/2r$ of non-overlapping 'neighbourhoods' or 'bins', which will turn out to be a relevant parameter for the collective dynamics. Finally, let $N$ denote the number of agents in this world.

In our model, the percept $s$ available to an agent is the net flow of individuals within range $r$ relative to the agent, that is, the number of neighbours going in the same direction as the agent minus those going in the opposite direction, distinguishing only absolute values up to two: $s \in \{-2, -1, 0, +1, +2\}$. In response to this percept, an agent can choose to either turn around ($a = -1$) or continue in the same direction ($a = +1$). Notice that the individuals take their actions in turns, so that, while one individual is deliberating, none of its neighbours will change direction. The reward scheme, for this first exploration of density-dependent alignment, is deliberately simplified: an agent is given a reward $R = 1$ if, at the end of its turn (i.e., after taking its action), it is moving in the same direction as the majority of its neighbours.

## 3 Density-dependent alignment of agents with fixed interaction rules

The first test of our agent-based model is whether agents with a single, fixed interaction rule can produce different regimes of collective alignment as a function of swarm density. To this end, we disable the learning subroutine in PS agents and instead fix the $h$-values which describe the percept-action mapping in their memories. In the simplified scenario outlined above, one can reasonably approximate an agent's responses with a single parameter, the decisiveness $d$, which controls how likely an agent is to follow the net movement of its neighbours. This can be represented by $h$-matrices of the form:[4]

$$\begin{aligned} h\left(s=+2, a=+1\right) = h\left(s=-2, a=-1\right) &= 1 + d \\ h\left(s=+1, a=+1\right) = h\left(s=-1, a=-1\right) &= 1 + \tfrac{1}{2}d \\ h\left(s, a\right) &= 1 \text{ otherwise.} \end{aligned} \qquad (3)$$

Fig. 1 shows the results of simulations wherein varying numbers of individuals with fixed interaction rules (fixed $d$) are placed in a world of fixed size. One can see that our model can account for the different regimes of collective dynamics as a function of density that are observed in experimental data. Reproducing this phenomenon is one of the most basic benchmarks for a model of locust motion, and achieving this puts our agent-based model on par with previous models, both individual-based and coarse-grained. However, we stress that our model does not require any fine-tuning of the individual-level interaction rules in order to account for the different regimes of motion, as some individual-based models do. In order to develop a more detailed quantitative analysis of the predicted collective dynamics, we will now introduce the additional formalism of the Fokker-Planck equation.

## 4 Comparing models in terms of the Fokker-Planck equation

### 4.1 Representing group-level dynamics

Plots of global alignment as a function of time, $z(t)$, such as those in Fig. 1, provide an intuitive visual representation of the most striking features of density-dependent alignment. However, each such plot only depicts a single possible evolution $z(t)$ and therefore fails to address important questions about the general features of the dynamics. We are particularly interested in features of the *collective* motion, i.e., features that persist as one coarse-grains over the detailed individual movement within the swarm, for instance how strongly aligned the metastable states are and how quickly the system transitions between them.

---

[4]Note that an agent whose memory is described by this $h$-matrix, when presented with the percept $s = 0$ (zero net flow of neighbours), may turn or continue with equal probabilities. This behaviour also emerges in a PS agent whose $h$-matrix is not fixed (see Fig. 3): unless some preference is acquired based on asymmetric rewards, all actions are taken with equal probability. In biological entities, on the other hand, not turning can be considered a default action, being energetically more favourable, and one would expect individuals to turn with probability less than $\tfrac{1}{2}$. This feature can be included in our model by introducing an 'inertia' parameter, which makes $h\left(s=0, a=+1\right) > h\left(s=2, a=-1\right)$, which entails $P\left(a=+1|s=0\right) > \tfrac{1}{2}$. Our simulations indicate that this modification has no discernible effect on the phenomena we describe in the following.

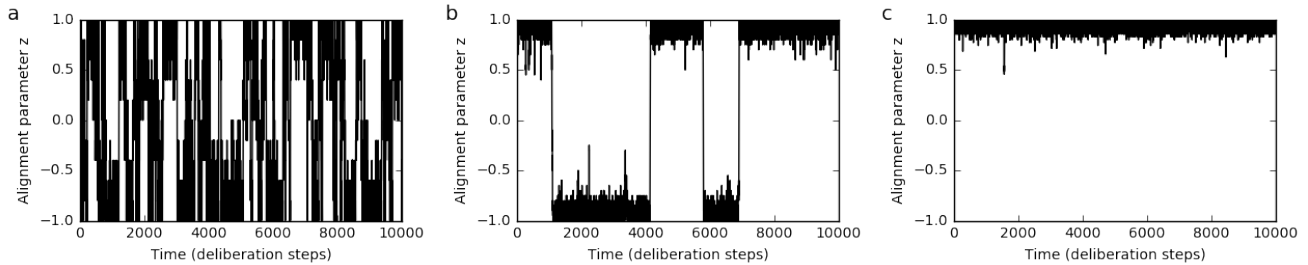

Figure 1: Different regimes of collective motion for agents with fixed decisiveness $d = 30$, as a function of density: (a) $N = 10$, (b) $N = 40$, (c) $N = 70$, for fixed $W = 80$, $r = 5$. Plots show the alignment parameter $z$ (group average of the direction of motion) as a function of time, with one time-step corresponding to one cycle of interaction and deliberation by the PS agent.

A useful tool for studying the coarse-grained limit is the Fokker-Plack equation (FPE), which describes the time evolution of a probability distribution over the alignment parameter, $P(z, t)$ – a natural choice for a system that is intrinsically stochastic. The FPE has been used to describe the alignment of marching locusts in previous work, e.g. by Dyson *et al.* [4] and by Ariel and Ayali [15], and proved a convenient representation for the purpose of comparing predictions of different models. It takes the form

$$\frac{\partial}{\partial t} P(z, t) = -\frac{\partial}{\partial z} [F(z) P(z, t)] + \frac{\partial^2}{\partial z^2} [D(z) P(z, t)]. \quad (4)$$

The coefficient $F(z)$ gives the drift velocity with which an element of probability $dP$ moves along $z$ in time, while $D(z)$ governs the diffusion of the probability density.

The functional dependence of the two coefficient functions on $z$ encodes essential features of the group-level dynamics. Most importantly, fixed points are given by the zeroes of the drift function; that is, by states with $z_0$ such that $F(z_0) = 0$, with $\frac{dF}{dz}(z_0) < 0$ indicating stability or equilibrium point, while $\frac{dF}{dz}(z_0) > 0$ implies unstable fixed points. Outside of fixed points, large-magnitude drift entails quicker transitions between states, while strong diffusion makes the system more likely to leave a given state.

One can show that the agent-based model we propose gives rise to a FPE when taking a suitable limit, as detailed in appendix A. In the next section, we will study the drift and diffusion functions that arise from this model and use them for a comparison with other results presented in the literature. More precisely, we will consider the drift function $F(z)$, which appears above, and the rescaled diffusion function

$$D'(z) \equiv N D(z). \quad (5)$$

The purpose of this rescaling is to compensate for the fact that, as the number of individuals $N$ increases (in the continuous limit, $N \to \infty$), the diffusion coefficient $D(z)$ vanishes as $1/N$. This $1/N$-decrease should not come as a surprise, since it is well-known that diffusion (or noise) becomes *proportionally* less relevant as the size of a system increases. The object of interest here, however, is how the diffusion varies *in addition* to this scaling, and in order to isolate this effect, it is preferable to analyse $D'(z)$. One can verify that, if one keeps the decisiveness $d$ and the density $N/B$ constant while varying the number of individuals, then $D'(z)$ remains unchanged. (Further details on this point are provided at the end of appendix A.)

## 4.2 Comparison of predictions in terms of the FPE

Fig. 2a,b shows the predictions of our agent-based model for the drift and diffusion coefficient functions of the FPE. We explore the effects of varying two parameters: the effective density $N/B$ and the decisiveness

$d$. For comparison, Fig. 2c reproduces the corresponding curves generated by a model fitted to experimental data by Dyson *et al.* [4], for different numbers of individuals $N$ confined to an arena of fixed size.

The essential features observed in the experimental data are reproduced by our model. At low densities, there exists a single stable state (with zero drift and relatively high diffusion): this is the completely disordered state, with the alignment parameter $z = 0$. At higher densities, two additional stable fixed points arise at $z \approx \pm 1$, which correspond to ordered collective motion in one direction or the other. At the same time, the disordered state becomes unstable, since drift tends to move the system away from $z = 0$.

Going beyond the effects of density, our model also allows us to explore the influence of a second degree of freedom, namely the decisiveness governing interactions at an individual level. To our knowledge, this degree of freedom was not considered explicitly in the experiments performed to date, but if one wishes to provide an account of collective dynamics as arising from individual behaviour, then some parameter(s) governing that behaviour must feature prominently in the model. The effects of the decisiveness parameter $d$ in our model are illustrated in Fig. 2b, and one can see that they are are qualitatively very similar to the effects of the density parameter discussed above. This implies that, while the different regimes of collective motion can be triggered by changes in density, they could equally well be due to changes in the interaction parameters at an individual level.

When comparing plots of $F(z)$ and $D(z)$ from different sources, one should note that the scaling of the ordinate axes reflects the scaling of the time variable in the FPE. In the experiments discussed and analysed in Fig. 2c, time is measured in steps of $0.8s$ and $0.2s$, respectively, for drift and diffusion. In the agent-based model, on the other hand, the natural unit is the deliberation time-step, that is, the time-scale on which individuals interact. If the relation between these two time-scales is unknown, then the predictions of the two models can only be meaningfully compared up to a scale factor.

The most striking difference between the experimental data reproduced in Fig. 2c and the predictions of our model concerns the scaling of the diffusion function with the number of individuals, which can be seen most easily at $z = 0$: in the results of Dyson *et al.*, $D(z = 0)$ grows approximately linearly with $N$, whereas in our model, the rescaled diffusion coefficient $D'(z) \equiv ND(z)$ at $z = 0$ is approximately independent of $N$. (A similar scaling with $N$ can be observed in the drift function $F(z)$, although assessing the scaling of a third-order polynomial is less straightforward.) However, it is unclear whether this difference in scaling factors is a symptom of an actual discrepancy or merely an artifact: as we pointed out above, the choice of time units directly affects the scaling of the drift and diffusion functions, and Dyson *et al.* do rescale their time variable by a factor of $N$ at one point, which could account for the differences observed here. We were unable to locate any other sources that present experimental results in a form that could be translated to FPE coefficients in order to clarify this point. In the absence of experimental data, one can at least compare our predictions to those of other theoretical models, for instance the models attributed to Czirók and Buhl that are disucssed by Ariel and Ayali (see [15] and references therein). They consider a diffusion coefficient in the limit of continuous $z$ – which corresponds to our $D'(z)$ – and find that its value at $z = 0$ is, to good approximation, independent of the density, just as it is in our model.

Leaving aside the question of the scale factor, one may note that the curves in Fig. 2a and c have some more subtle differences in shape, such as the relative height of the extrema $F(z = \pm 1)$ compared to the local extrema at intermediate $z$. When considering these differences, one should bear in mind that the curves reproduced in Fig. 2c were constructed by fitting experimental data to a particular theoretical model, which, due to the form of the interactions that is assumed, constrains $F(z)$ and $D(z)$ to be third-order polynomials. In particular, the relatively large absolute values of $F(z = \pm 1)$ in Fig. 2c should be considered mostly an extrapolation from the data for other values of $z$, since the experimental system rarely visits states with $|z| \to 1$ and consequently experimental data for that interval is scarce and subject to considerable statistical uncertainty.

The representation in terms of the FPE can also be used to compare the predictions of different theoretical models. This was done by Ariel and Ayali [15] for three different models and varying densities. The predictions of two of the models (attributed to Czirók and Buhl) are quite similar to each other and can be accounted for by our model by taking suitable values of the parameters. Meanwhile, the pause-and-go model [7] makes some qualitatively different predictions, such as the existence of a *stable* fixed point at $z = 0$ even at high densities. These differences can be attributed to the distinguishing feature of the pause-and-go model, namely that it allows individuals to stand still temporarily. We stress that, while the particular model considered here does not allow for pausing and consequently does not predict a stable fixed point at $z = 0$, the explicitly agent-centric formulation of the class of models that we are proposing makes it straightforward to include this feature if desired. More generally, agent-based models offer the possibility of incorporating a wide variety of features of individual interactions by adapting the spaces of percepts and actions to reflect the actual animal's sensory and motor capabilities.

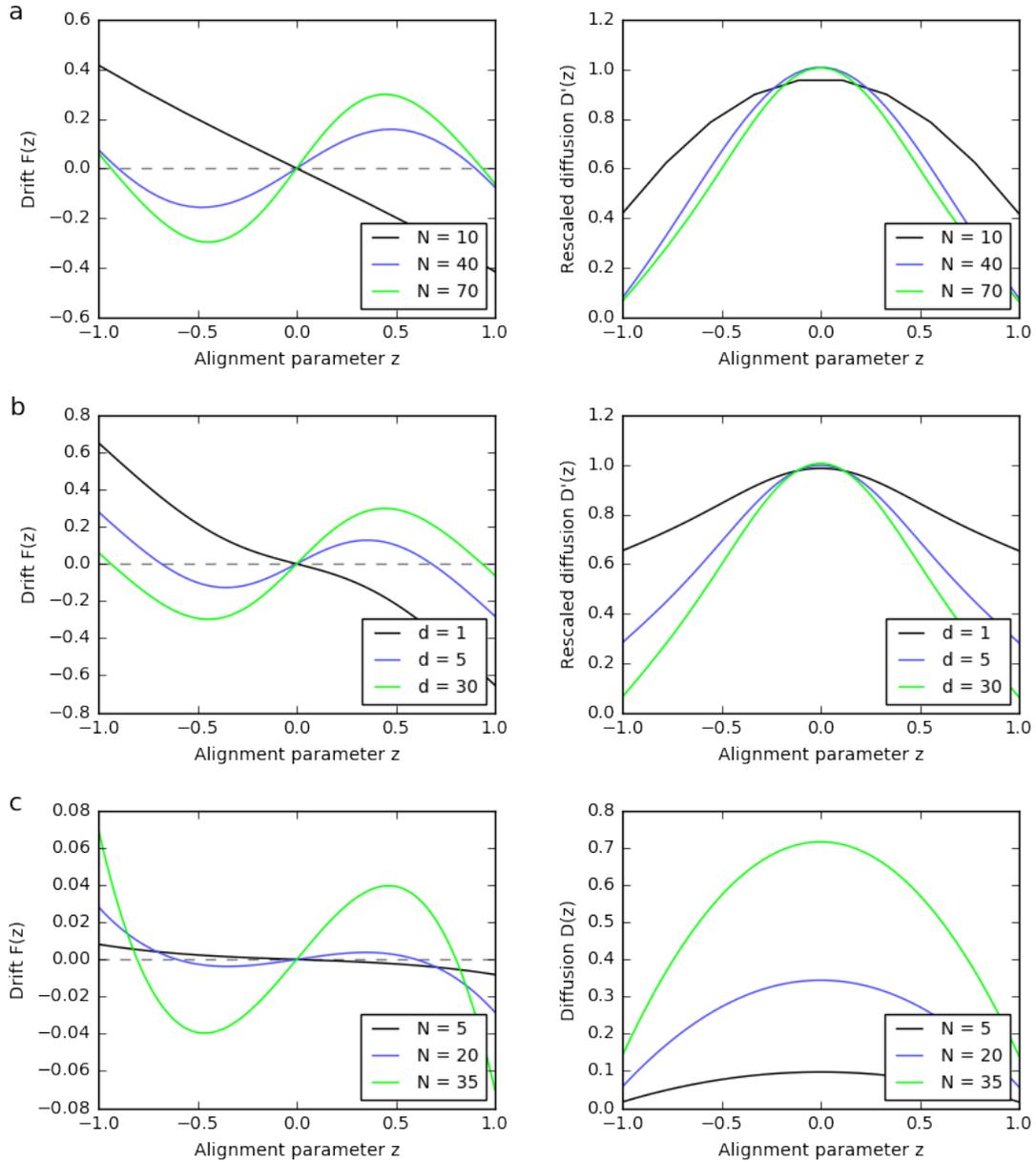

Figure 2: Group-level dynamics, described by the two coefficient functions in the FPE; drift $F(z)$ (left) and rescaled diffusion $D'(z)$ (right). We show (a) predictions of an agent-based model with fixed decisiveness $d = 30$ and variable effective density $N/B$ (number of individuals per neighbourhood), fixing $B = 8$, (b) predictions with fixed effective density ($N = 70$, $B = 8$) and variable decisiveness. Panel (c) reproduces the coefficient functions for drift $F(z)$ (left) and diffusion $D(z)$ (right) of Dyson *et al.* [4], which were obtained by fitting experimental data to a model based on transition rates, for different numbers of individuals $N$.

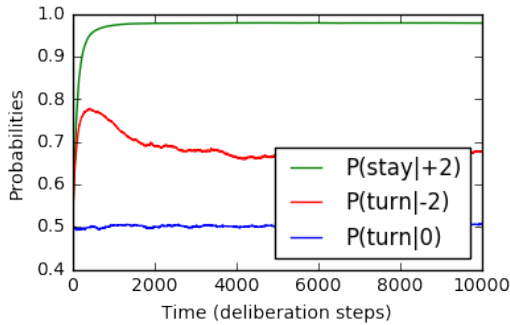

Figure 3: Learning individual responses: probabilities of various responses to percepts as a function of time (measured in discrete interaction steps), averaged over all agents, for $N = 40$, $W = 80$, $r = 5$ ($B = 8$) and $\gamma = 0.002$. Curves depict the probabilities of individuals (green) maintaining their direction when perceiving a net flow in the same direction they are moving, (red) turning around when facing a net flow in the opposite direction, and (blue) turning around when perceiving zero net flow.

## 5 Learning individual-level interaction rules

In the previous two sections, we have argued that agent-based models offer a more detailed and potentially more realistic account of the perceptions, responses and internal processes of individuals in a swarm than traditional individual-based models. We now turn to the second major advantage of agent-based models: the fact that interaction rules need not be postulated, but can instead be learned. In the case of PS agents, as introduced in section 2.1, these rules are encoded in the $h$-matrix, which describes how the agent responds to different percepts. For the purpose of a rough comparison in the one-dimensional alignment task, these responses can be summarized by three probabilities: that of turning around when perceiving strong net flow in the opposite direction ($s = -2$), that of maintaining the direction when perceiving strong net flow in the same direction ($s = +2$), and that of turning given zero net flow ($s = 0$). We expect that a 'sensible' agent would have high probabilities of the first two, while the response to $s = 0$ is random, with probability $\frac{1}{2}$. Fig. 3 shows that our learning agents do develop these responses given the feedback from the environment. Moreover, the learning process proceeds quite quickly, over time-scales of the order of $10^2$ time-steps, which is not unexpected given the small spaces of percepts and actions in our model.

One may note in Fig. 3 that learning agents are more likely to maintain their direction given the percept $s = +2$ than they are to turn around given the percept $s = -2$. This is due to a self-reinforcing statistical effect: as an agent becomes more likely to align itself with the majority of its neighbours, percepts indicating a net flow in the opposite direction are encountered less frequently. Consequently, the asymptotic $h$-values, which are determined by the balance of reinforcement and forgetting, are smaller than for more frequently encountered percepts. At the same time, they are subject to larger statistical fluctuations, as one can also see in Fig. 3. Meanwhile, the percepts representing a net flow in the *same* direction as the agent are received increasingly frequently, so that the $h$-value associated with the correct response (namely to keep going in the same direction) is reinforced more often. It follows that the decisiveness with which the agent responds to these two types of percepts becomes markedly different with time. Despite this deviation from the fixed-interaction form assumed in section 3 (specifically eq. (3)), the collective motion of learning agents also reproduces the phenomenon of density-dependent alignment, as can be seen in Fig. 4 and by the drift and diffusion functions in Fig. 5.

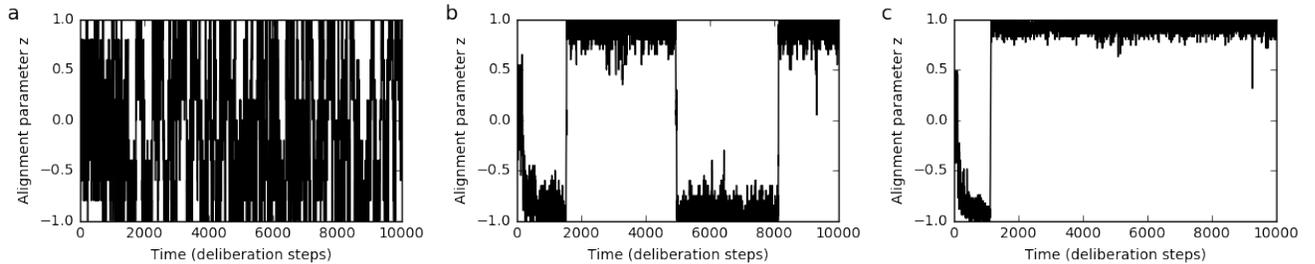

Figure 4: Global alignment $z$ over time for learning agents: (a) low density ($N = 10$) leads to disordered motion, (b) intermediate density ($N = 40$) leads to alignment with periodic switching, and (c) high density ($N = 70$) leads to long-term alignment. The appearance of well-defined regimes of collective dynamics is preceded by a transient period, during which agents' individual behaviour is still changing. Parameters: $W = 80$, $r = 5$, $\gamma = 0.002$.

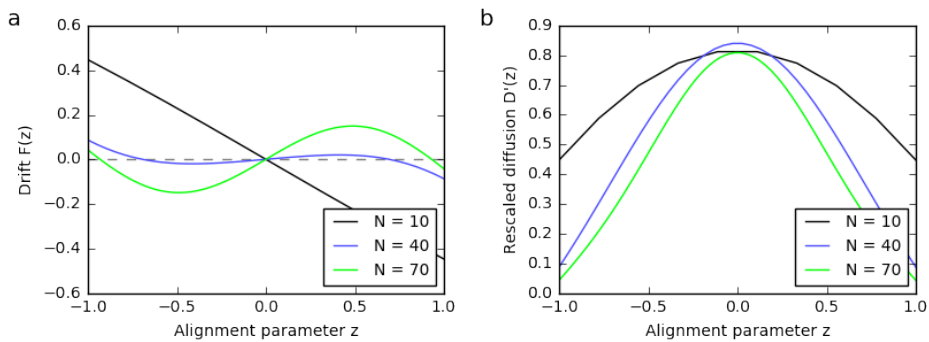

Figure 5: Group-level dynamics of learning agents, described by the two coefficient functions in the FPE: (a) drift $F(z)$ and (b) rescaled diffusion $D'(z)$. The group dynamics are based on the $h$-matrix developed by learning agents after $10^4$ interaction steps, averaged over the ensemble of agents. Parameters: $W = 80$, $r = 8$, $\gamma = 0.002$.

One can also see in Fig. 4 that our model captures a transient regime, which arises while the individual-level behavioural dispositions (technically, the response probabilities derived from the $h$-matrix) are still changing noticeably. Such transient changes of invididual-level behaviour could serve as a model of the changes undergone by real locusts as they transition from the solitarious to the gregarious phase. Given the crucial role that this transition plays in the formation of swarms and plagues, having a theoretical model for predicting collective dynamics which naturally accommodates changing individual behaviour could be extremely useful.

Finally, since the PS model explicitly includes noise in the learning process, in the form of the forgetting parameter $\gamma$, it also allows one to consider how the learning process would turn out for agents that do not forget, with $\gamma = 0$. It is easy to see that, in the limit of long learning times, these agents will develop *deterministic* responses to their neighbours' movement, as the $h$-values associated with rewarded actions grow without bound. This is not very realistic, since the processes that govern the movement of real animals are generally noisy – to a greater or lesser extent –, making the behaviour stochastic. In fact, some measure of unpredictability was shown to be advantageous for survival, for instance when evading predation [16].

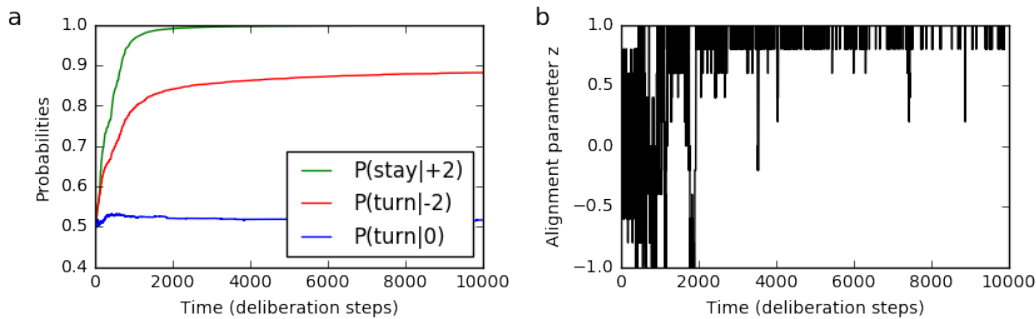

Figure 6: Agents that learn without forgetting ($\gamma = 0$): (a) Response probabilities grow towards unity. (b) The collective motion tends to be strongly aligned, with $z(t)$ fixed at $\pm 1$ for long times. This is noteworthy since the scenario presented here has relatively low density, with the same parameter values for $N$ and $B$ that resulted in completely disodered motion in agents that do forget (cf. Fig. 4a). As one moves to higher densities, the tendency towards long-term collective alignment becomes even more pronounced. Parameters: $N = 10$, $W = 80$, $r = 5$ ($B = 8$).

Moreover, a deterministic model could not account for the observed collective motion, since – as discussed in the previous section – high decisiveness leads to long-term alignment even in the limit of low swarm densities. Both of these effects are illustrated in Fig. 6.

## 6 Discussion

The results detailed in the previous sections show that our agent-based model can reproduce a key qualitative feature of collective motion which has been observed in experiment, namely the density-dependent switching of group alignment in marching locusts. At a more quantitative level, our model can also account for the functional form of the drift and diffusion coefficients in the FPE governing global alignment. Indeed, one can show that (see appendix), by taking a suitable limit, agent-based models formally reduce to individual-based models that had been previously proposed to account for experimental data. However, when compared with the first generation of models, the agent-based ansatz offers several advantages, which we will explore in the following.

Firstly, as in other individual-based models, we aim to provide an account of the collective dynamics which is derived from a reasonable description of individual behaviour. This ansatz offers a more detailed and realistic understanding of the origin of collective behaviour than coarse-grained models can. As an illustrative example, consider the theoretical model of [4]: this model reduces the swarm to two populations, moving clockwise and counter-clockwise, respectively, and takes as fundamental parameters the transition rates between these populations. The model ignores the fact that these rates are only a coarse-grained description of the effects of distinct individuals receiving various stimuli and responding to them according to their intrinsic dispositions. Consequently, the model cannot predict whether or how these parameters must change with the number of individuals, $N$, which is the parameter that normally changes in the pertinent experiments. Upon fitting experimental data to this model, the transition rates are found to depend strongly (roughly linearly) on $N$. These types of effect can only be predicted and understood by delving into the details of how the individual members of the swarm interact.

Within the family of individual-based models, one is generally free to assume that the rules according to which the individuals interact take an arbitrary mathematical form – possibly changing sharply depending on some condition – and that they depend on any variables that are represented in the model. By contrast, agent-based models are designed to provide a more natural and realistic account of which perceptions individuals respond to and how these are processed to produce a response. For example, the model presented here introduces a parameter that measures how likely each individual is to respond to the movement of their neighbours, rather than imposing that the individual's speed is deterministically set to some weighted local average broadened by an additive noise term. Using this model, we have shown that a single individual-level interaction rule can account for the different regimes of collective motion that are observed at different densities (see section 3). Notably, there is no need to adjust the value of the decisiveness parameter as a function of density. This feature of our model matches the empirical finding that (once the initial process of behavioural gregarization is complete) the locusts' activity at the individual level remains fixed across "a wide range of densities" [3].

Furthermore, while it is possible to account for the different regimes of collective alignment by varying the density alone, our model also offers the possibility of changing the parameters governing individual behaviour in order to explore their effect on the collective motion. This possibility is particularly intriguing in the context of the transition from the solitarious to the gregarious phase in locusts and the role that this change of intrinsic characteristics of the individuals may play in the formation of swarms. It would be very interesting to test this model against experimental observations of locusts in different behavioural phases, which could be obtained, for instance, by exploiting the fact that behavioural changes take some time (on the order of hours [17]) to come into full effect when locusts are crowded.

On the topic of experimental implementations, we note that a move towards more detailed models of the behaviour of individual locusts is particularly timely in light of the advances in data acquisition and analysis that were made in the last decade. In particular, the methods for automated tracking and analysis of the movement of individual locusts introduced in [3, 18] provide a natural experimental tie-in for agent-based models.

While agent-based models provide a more fine-grained understanding of the origins of collective dynamics, one can also show that they reduce to previously established, more coarse-grained models in a suitable limit. This makes it easy to fit this new class of model into existing data-analysis frameworks and, in particular, compare predictions to previously obtained results (both theoretical and experimental). We have illustrated this procedure in section 4 by deriving a Fokker-Planck equation for the case of locusts marching in a one-dimensional arena. This form has been used to compare the predictions of different models in previous literature [15].

A second major advantage of agent-based models is the possibility of agents learning the interaction rules based on feedback from the environment, rather than being programmed with them *ad hoc*. The results presented in section 5 show that agents which are rewarded for aligning themselves with their neighbours learn to respond to their neighbours' movements in a way that reproduces density-dependent alignment at the swarm level. In a similar way, agents which are presented with a richer perceptual space can autonomously learn which percepts are relevant to them. This avenue of research should also provide predictions that can be tested against feasible experiments and allow us to further our understanding of the mechanisms which lead to the formation of swarms.

The ability to learn, and the notion of agency which it supports, merit some comment. In a typical case of reinforcement learning (in the context of artificial intelligence), the learner is a single individual, which gradually develops particular behavioural tendencies based on its history of interactions with the environment. Since these interactions – and ultimately the behaviours that they give rise to – are specific to a particular individual, one can argue that the individual has become an *agent*: acting on its own, in the sense that its behaviour is governed by factors that are unique and intrinsic to the individual. However, in the process that we model here, namely the development of collective behaviour in animals, the adaptation takes place over many generations, with little change within the lifetime of a single individual. In particular, in the case of locusts, single individuals only exhibit very limited behavioural plasticity over the course of their lives.[5] Consequently, such simple animals cannot be considered agents in the above sense, being instead better described as automata that carry out a fixed set of instructions. However, the same is not true for the individuals in our simulations, which can be interpreted as avatars of entire generations. In this sense, the property of agency is recovered as one moves to the longer time-scales of evolutionary adaptation. In the case of more complex animals, on the other hand, which are capable of meaningfully modifying their behaviour over the course of a single individual's lifetime, the learning process of a simulated agent may serve as a direct model of the learning process of actual animals. Considering the versatility of the full PS framework (which was demonstrated for a range of scenarios in [20, 21, 22, 23]), this is a promising direction for further work.

---

[5]The most striking example is the switch between the solitarious and gregarious 'programmes', although more recently some evidence of associative learning was also observed [19].

Since the possibility of learning interaction rules is a central feature of agent-based models, the reward scheme which drives this learning process plays an important role. While it may seem more involved, at first sight, to specify what locusts are rewarded for instead of describing directly how they behave, we argue that our reinforcement-based ansatz offers greater insights, since it includes the reasons *why* individuals behave as they do. In the present work, one may note that the learning task faced by the agents is rather basic: given a percept that encodes the net movement in the vicinity, an agent is given the choice of either turning around or not and is subsequently rewarded if it successfully aligned itself with its neighbours. Ideally, one should specify a more realistic condition for being rewarded (such as first collecting a large group and subsequently traversing long distances under adverse conditions) in order to test whether cohesion and density-dependent alignment emerge spontaneously from these requirements. While such a problem would make for a more demanding learning task, due to the necessarily larger percept space and highly delayed rewards, this is a natural direction for the next steps in our research. The simpler task considered here was intended only to illustrate that learning agents can improve our understanding of the origins of collective motion *in principle*. In light of the results discussed above, we now look forward to combining our ansatz with more realistic survival tasks, based on insights from biology and ecology, in order to explore the phenomenon of collective motion in more complex settings.

**Acknowledgements:** This work was supported in part the Austrian Science Fund (FWF) through the SFB FoQuS F4012, and by the Ministerium für Wissenschaft, Forschung, und Kunst Baden-Württemberg (AZ: 33-7533.-30-10/41/1).

# A From learning agents to a Fokker-Planck equation

This appendix begins with the agent-based model presented in the main paper and derives a continuous-variable description of the group dynamics, in the form of a Fokker-Planck equation for the probability density $P(z,t)$. The derivation proceeds as follows: We begin by focusing on a single (generic) individual and computing the probabilities that this individual receives different percepts, given the current value of the global alignment parameter. Combining this information with the conditional probabilities for turning or continuing in the same direction, which are derived from the $h$-matrix of the PS model, one can then compute the probability that a given individual will turn around in the current time-step. From this, in turn, one can in general compute how many individuals convert from going in one direction to the opposite, and consequently the global alignment of the population in the next time-step. Since the entire process is stochastic, this yields a probability distribution over $z$ at the next time-step. Taking the limit in which the time $t$ and the alignment parameter $z$ become continuous, the dynamics can be cast as a Fokker-Planck equation, whose drift and diffusion coefficients we derive as a function of – ultimately – the $h$-matrix, which describes the memory of the individual locust. This form will allow us to compare the predictions of our model to the results of other works.

## A.1 Derivation of group-level transition probabilities

We begin by establishing the relevant variables. In addition to the total number of agents $N$, we will use the combination $B = W/2r$, which specifies how many neighbourhoods (or 'bins', defined as regions that an individual can see) the world is divided into. For the first part of the derivation, which is cast explicitly in terms of discrete individuals, it is convenient to use the variable $X_+$ denoting the number of individuals moving in the clockwise direction (which we arbitrarily label as positive) at the beginning of the current time-step, while $X_- = N - X_+$ individuals are moving anticlockwise. This information can equivalently be expressed by the alignment parameter $z = \frac{1}{N}(X_+ - X_-)$.

Focusing on a single, generic agent, the *focal agent*, let $\tilde{X}_\pm$ denote the numbers of individuals moving in each direction within this agent's sensory range, while $\epsilon = \pm 1$ indicates in which direction the agent itself is moving. The percept received by this agent is then

$$s = \epsilon \left( \tilde{X}_+ - \tilde{X}_- \right) \in \{-2, -1, 0, +1, +2\}, \tag{6}$$

truncated to absolute values of at most 2. Based on the probabilities of a given individual turning around, we will compute the number of individuals turning from positive to negative (clockwise to anticlockwise), denoted $D_-$, and the number turning the opposite way, denoted $D_+$. The variable of interest in this calculation is the number of individuals moving clockwise in the subsequent time-step, $X'_+ = X_+ + (D_+ - D_-)$, or equivalently the change $\Delta X_+ = D_+ - D_-$.

**Probability distribution over percepts.** We begin by determining the probability that the focal agent receives a certain percept, which is essentially the difference between the numbers of individuals moving each way within the its sensory range, $\tilde{X}_+ - \tilde{X}_-$. In order to derive a tractable expression for this quantity, we make three assumptions:

- In order to derive transition probabilities that depend only on the fraction of individuals going each way but not on their individual positions, we will assume that the individuals are approximately homogeneously distributed in space.

- Moreover, we assume that individuals are *independently* distributed in space. (This assumption is inaccurate in the limit of low densities, when, according to our simulations, individuals tend to congregate in groups with density $\approx 1/r$. However, as soon as the number of agents relative to the size of the world is high enough, $N/W > 1/r$, it becomes reasonable to assume a homogeneous, independent distribution.)

- Finally, we will neglect the fact that the overall number of individuals going in each way, $X_\pm$, is in fact finite. This is justified by different considerations, depending on the regime: If $B \gg 1$, then only a small fraction of the total $X_\pm$ is located within a given bin. Therefore, for the purpose of determining how many individuals are in this bin, the approximation that one is drawing from an infinite pool is reasonable. If, however, $B \to 1$, then the number $\tilde{X}_\pm$ of individuals moving in a given direction within the focal agent's sensory range is most likely large anyway (of order $N$), and since percepts only distinguish $\tilde{X}_+ - \tilde{X}_-$ up to absolute values of 2, overestimating these numbers is unlikely to cause deviations.

Under these assumptions, the numbers $\tilde{X}_\pm$ follow a Poisson distribution with mean $X_\pm/B$. (One may note that the means of the distributions $P\left(\tilde{X}_\pm\right)$ are related by the constraint $X_+ + X_- = N$. However, the particular values for $\tilde{X}_\pm$ that we draw from these distributions – that is, how many of the $X_\pm$ individuals going each way are within the focal agent's sensory range – are statistically independent.[6]) The difference between two Poisson-distributed variables with means $\mu_{1,2}$ follows a Skellam distribution,

$$\text{Skellam}(s'; \mu_1, \mu_2) \equiv e^{-(\mu_1 + \mu_2)} \left(\frac{\mu_1}{\mu_2}\right)^{\frac{s'}{2}} I_{s'}\left(2\sqrt{\mu_1 \mu_2}\right), \quad (7)$$

where $I_{s'}(z)$ denotes the modified Bessel function of the first kind. In our case, this gives the probability distribution over $s'$ conditioned on (i.e., if one knows the value of) the total number of individuals moving in the positive direction, $X_+$:

$$P\left(\tilde{X}_+ - \tilde{X}_- = s'|X_+\right) = \text{Skellam}\left(s'; \frac{X_+}{B}, \frac{X_-}{B}\right) = e^{-N/B} \left(\frac{X_+}{X_-}\right)^{\frac{s'}{2}} I_{s'}\left(\frac{2}{B}\sqrt{X_+ X_-}\right), \quad (8)$$

where one can replace $X_- = N - X_+$. (Note that we condition only on $X_+$, since including $X_-$ in the known information would be redundant.)

Notice that the difference $\tilde{X}_+ - \tilde{X}_- = s'$ can in principle run over all integers; only in the agent's perception are values $s' \geq 2$ resp. $s' \leq -2$ combined into a single percept each. The sum

$$P\left(\tilde{X}_+ - \tilde{X}_- \geq 2|X_+\right) = e^{-N/B} \sum_{s'=2}^{+\infty} \left(\frac{X_+}{N - X_+}\right)^{\frac{s'}{2}} I_{s'}\left(2\sqrt{X_+(N - X_+)}\right), \quad (9)$$

is simply an element of the cumulative density function associated with the Skellam distribution, and analogously $P\left(\tilde{X}_+ - \tilde{X}_- \leq -2|X_+\right)$. If the focal agent is currently moving in the positive (clockwise) direction, then this probability distribution is precisely the probability distribution over percepts, $s = s'$ (with the cutoff $|s| \leq 2$), whereas for agents moving in the negative direction, the percept is $s = \tilde{X}_- - \tilde{X}_+ = -s'$ (again enforcing $|s| \leq 2$).

**Probability of turning.** In the PS model, the $h$-matrix determines the probabilities of turning around given a percept, $P(turn|s)$. Together with the probabilities computed in the previous section, $P(s|X_+, \epsilon)$, this allows us to obtain

$$P(turn|X_+, \epsilon = \pm) = \sum_s P(turn|s) P(s|X_+, \epsilon), \quad (10)$$

which indicate how likely it is that a particular individual from either sub-population (clockwise or anticlockwise) turn around in a given time-step. Notice that, under the assumptions detailed above, this probability depends only on the global variable $X_+$ and the individual's orientation $\epsilon$.

---

[6] In fact, this statistical independence also holds for finite $N$, in which case $\tilde{X}_\pm$ follow binomial distributions.

**Probability distribution over populations $X'_\pm$.**   In order to obtain the populations going in each direction at the next time-step, $X'_\pm$, we will now compute the numbers of individuals turning from the positive to the negative direction (clockwise to anticlockwise), denoted $D_-$, and the number turning from negative to positive, denoted $D_+$. Notice that only the difference $D_+ - D_-$ manifests as an effective increase of the population $X_+$, but for the purpose of computing transition probabilities, one must distinguish how many individuals turned in each direction, even if their numbers partially cancel afterwards.

Since the individuals that turn, $D_\pm$, are necessarily a subset of those currently moving in the direction in question, $X_\mp$, we model $D_\pm$ as following a binomial distribution:

$$P(D_\pm | X_+) = \text{Binom}(D_\pm; X_\mp, p = Prob(turn|\mp)) = \binom{X_\mp}{D_\pm} p^{D_\pm} (1-p)^{(X_\mp - D_\pm)}. \qquad (11)$$

The probability distribution over $\Delta X_+ = D_+ - D_-$ resulting from this model does not have a closed form, requiring instead a sum over the various combinations of $D_\pm$ that lead to each $\Delta X_+$. Specifically, letting $\mathcal{D}$ denote the set of ordered pairs $(D_+, D_-)$ such that $D_+, D_- \geq 0$ and $\Delta X_+ = D_+ - D_-$, we can write

$$P(\Delta X_+ | X_+) = \sum_{(D_+, D_-) \in \mathcal{D}} P(D_+ | X_+) P(D_- | X_+). \qquad (12)$$

**Result: discrete-time transition matrices.**   The above method allows one to derive a stochastic rule that describes the changes to the probability distribution over the alignment parameter. In the discrete case, $P(X_+)$ can be represented as a vector, and the transition probabilities $P(X'_+ | X_+)$ take the form of a matrix. In the limit of large population size $N$, it is convenient to replace the argument $X_+$ by $z = \frac{1}{N}(X_+ - X_-)$, which becomes continuous as $N \to \infty$, making the distribution $P(z)$ and the conditional $P(z'|z)$ functions of one resp. two continuous variables, with range $[-1, 1]$. The conditional probability $P(z'|z)$ specifies how the probability distribution $P(z)$ changes in a time-step $\Delta t$ (that is, during the time it takes for an agent to deliberate and choose its next action), and it allows one to read off key features of the collective dynamics, such as whether there are metastable states, how strongly aligned the group is in these states, and how quickly the system transitions between them.

The transition functions generated by our model, with the fixed $h$-matrix given by eq. (3), are shown in Fig. 7. All instances exhibit a narrow band of non-negligible probabilities, which implies that the mapping from $z$ at time $t$ to $z'$ at time $t + \Delta t$ is approximately deterministic. If this band lies in the diagonal $z' = z$, then the alignment parameter tends to remain unchanged at any value. Fig. 7 shows how the dynamics deviates from this default in response to two parameters: Firstly, as the effective density $N/B$ increases, the peak of $P(z'|z)$ remains at a fixed, large $|z'|$ for a wider range of $z$. That is, the dynamics maps a wider range of intermediate states $z$ to a particular pair of strongly aligned states. This can be understood as a consequence of high densities quickly suppressing non-aligned states. Secondly, as decisiveness $d$ increases, the value of $|z'|$ to which the system tends increases; that is, the two metastable states between which the system alternates become more strongly aligned. This can be attributed to the high decisiveness making individuals less likely to turn against the majority. (One can see this effect clearly in the following example: if the group was initially perfectly aligned, $z = 1$, then the expected alignment at the next time-step is $z' = 1 - \frac{1}{1+d/2}$.)

## A.2   Making the alignment parameter $z$ and the time $t$ continuous

**The limit of continuous time and transition rates.**   The model derived above gives the probabilities of finite changes in the populations $X_\pm$ over discrete and finite time-steps $\Delta t$, which is natural in the context of reinforcement learning. However, in order to relate our work to other models that may not necessarily assume discrete time, we will now modify the above treatment to recover continuous time. To this end, we will introduce an infinitesimally small interval $\delta t$ and determine the transition probabilities $P_{\delta t}(X'_+ | X_+)$ for this time-step.

In order to derive group-level transition probabilities $P_{\delta t}(X'_+ | X_+)$ for an infinitesimally small $\delta t$, we begin with the following consideration: if a single individual has a probability $P_{\Delta t}(turn|s)$ of turning around in a finite time interval of default duration $\Delta t$, then the probability of turning in a smaller time interval $\delta t$ is proportionately smaller, $\frac{\delta t}{\Delta t} P_{\Delta t}(turn|s)$. Formally, this assertion can be derived from the assumption that 'turning' is an instantaneous event that could happen with uniform probability at any time. The probability of not turning, which is the absence of such an event, is consequently $1 - P_{\Delta t}(turn|s)$ and $1 - \frac{\delta t}{\Delta t} P_{\Delta t}(turn|s)$, respectively. This prescription changes the probabilities (10) of turning and staying that we derive from the $h$-matrix (which is based on the finite $\Delta t$), giving

$$P_{\delta t}(turn|X_+, \epsilon = \pm) = \sum_s \frac{\delta t}{\Delta t} P_{\Delta t}(turn|s) P(s|X_+, \epsilon). \tag{13}$$

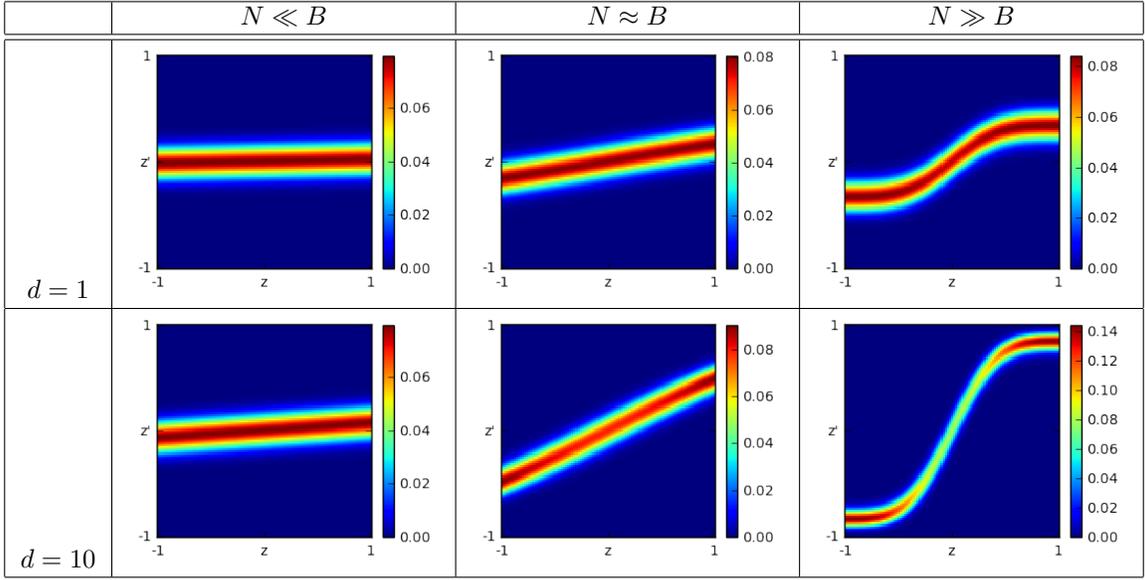

Figure 7: Stochastic transition functions $P(z'|z)$ describing the evolution of (the probability distribution over) the global alignment parameter, for the agents with fixed interaction rules described in eq. (3). Parameters: $N = 100$, $B \in \{1000, 100, 10\}$, $d \in \{1, 10\}$. (Note that the independent parameters $W$ and $r$ do not appear in the calculations leading to these plots; only the ratio $B = W/2r$ matters.)

It will be convenient to define transition *rates* (probability per time) for a given individual in a particular sub-population turning around, given the current value of $X_+$:

$$\tau_\pm(X_+) \equiv \lim_{\delta t \to 0} \frac{P_{\delta t}(turn|X_+, \epsilon = \mp)}{\delta t} = \sum_s \frac{P_{\Delta t}(turn|s)}{\Delta t} P(s|X_+, \epsilon = \mp). \tag{14}$$

This allows us to write the probabilities as

$$P_{\delta t}(turn|X_+, \epsilon = \pm) = \delta t \cdot \tau_\mp(X_+), \tag{15}$$

thereby making the dependence on the time interval explicit.

The remainder of the derivation of $P(\Delta X_+|X_+)$ proceeds as before, resulting in a combination of two binomial distributions. For notational simplicity, let us restrict ourselves to the case $\Delta X_+ \geq 0$. (The alternative case follows by exchanging $+$ and $-$.) Recalling that $\mathcal{D}$ is the set of $(D_+, D_-)$ such that $D_+, D_- \geq 0$ and $\Delta X_+ = D_+ - D_-$, this implies that we must sum over $D_+ \geq \Delta X_+$. On the other hand, since the individuals turning into the positive direction are drawn from $X_-$, it holds that $D_+ \leq X_-$. This gives

$$\begin{aligned} P_{\delta t}(\Delta X_+|X_+) &= \sum_{D_+ = \Delta X_+}^{X_-} \binom{X_-}{D_+} \binom{X_+}{D_+ - \Delta X_+} \\ &\quad \cdot [\delta t \tau_+]^{D_+} [1 - \delta t \tau_+]^{(X_- - D_+)} [\delta t \tau_-]^{(D_+ - \Delta X_+)} [1 - \delta t \tau_-]^{(X_+ - D_+ + \Delta X_+)}. \end{aligned} \tag{16}$$

As soon as $\delta t$ becomes small enough that $\tau_\pm \delta t \ll 1$ – in other words, that the turning probabilities for a single individual become small –, one can neglect the terms with large numbers $D_+$ and $D_- = D_+ - \Delta X_+$ of individuals turning in both directions. Assuming that this makes $D_\pm \ll X_\mp$, one can then approximate the binomial coefficients as $X_\pm^{D_\mp}/(D_\mp!)$, yielding

$$P_{\delta t}\left(\Delta X_+ | X_+\right) \approx \sum_{D_+ \geq \Delta X_+} \frac{1}{D_+!(D_+ - \Delta X_+)!} \left[X_- \delta t \tau_+\right]^{D_+} \left[X_+ \delta t \tau_-\right]^{(D_+ - \Delta X_+)}. \tag{17}$$

If one suppresses all but the lowest order in $X_\mp \delta t \tau_\pm$, which means neglecting all terms except $D_+ = \Delta X_+$, one obtains

$$P_{\delta t}\left(\Delta X_+ | X_+\right) \approx \begin{cases} \frac{1}{\Delta X_+!} \left[X_- \delta t \tau_+\right]^{\Delta X_+} & \Delta X_+ \geq 0 \\ \frac{1}{\Delta X_+!} \left[X_+ \delta t \tau_-\right]^{\Delta X_+} & \Delta X_+ \leq 0 \end{cases}. \tag{18}$$

One can see that the conditional probability $P_{\delta t}\left(X'_+ | X_+\right)$ becomes sharply peaked around $X'_+ = X_+$, with small values for $X'_+ = X_+ \pm 1$ and negligible values outside that region. That is, the most relevant quantities are

$$P_{\delta t}\left(X'_+ = X_+ \pm 1 | X_+\right) \approx \delta t \cdot X_\mp \tau_\pm (X_+). \tag{19}$$

Again, for notational simplicity, we introduce the transition rates

$$T_\pm(X_+) \equiv \lim_{\delta t \to 0} \frac{P_{\delta t}\left(X'_+ = X_+ \pm 1 | X_+\right)}{\delta t}. \tag{20}$$

Substituting the approximate expression for $P_{\delta t}\left(X'_+ = X_+ \pm 1 | X_+\right)$,

$$T_\pm(X_+) \approx X_\mp \tau_\pm (X_+). \tag{21}$$

As one should expect, the probability of any one out of $X_\mp$ individuals turning around grows linearly as one increases the number of individuals one is sampling from. This observation will become relevant in the next section.

**The limit of continuous alignment parameter $z$.** The final step is to make the alignment parameter $z \equiv \frac{2X_+}{N} - 1$ continuous, by letting $N \to \infty$. To this end, notice that one can rewrite the transition rates obtained in the previous subsection as

$$T_\pm(z) \equiv \lim_{\delta t \to 0} \frac{P_{\delta t}\left(z' = z \pm \frac{2}{N} | z\right)}{\delta t}. \tag{22}$$

In the limit of continuous time, as simultaneous transitions of more than one individual become negligibly unlikely, the balance of probabilities of (a) leaving the state with a particular $z$ and (b) reaching that state starting from nearby states with $z' = z \pm \frac{2}{N}$ is

$$\frac{\partial}{\partial t} P(z,t) = -\left[T_+(z) + T_-(z)\right] P(z,t) + T_+\left(z - \frac{2}{N}\right) P\left(z - \frac{2}{N}, t\right) + T_-\left(z + \frac{2}{N}\right) P\left(z + \frac{2}{N}, t\right). \tag{23}$$

One can now use a Taylor expansion around $z$ to rewrite this as

$$\frac{\partial}{\partial t} P(z,t) = -\frac{2}{N} \frac{\partial}{\partial z} \left[\{T_+(z) - T_-(z)\} P(z,t)\right] + \frac{4}{2N^2} \frac{\partial^2}{\partial z^2} \left[\{T_+(z) + T_-(z)\} P(z,t)\right] + \mathcal{O}\left(\frac{1}{N^3}\right). \tag{24}$$

This has the form of a Fokker-Planck equation,

$$\frac{\partial}{\partial t} P(z,t) = -\frac{\partial}{\partial z} \left[F(z) P(z,t)\right] + \frac{\partial^2}{\partial z^2} \left[D(z) P(z,t)\right], \tag{25}$$

if one identifies the drift and diffusion coefficients, respectively, as

$$\begin{cases} F(z) = \frac{2}{N}\left[T_+(z) - T_-(z)\right] \\ D(z) = \frac{4}{2N^2}\left[T_+(z) + T_-(z)\right]. \end{cases} \qquad (26)$$

It is instructive to rewrite these expressions in terms of more fundamental quantities, in particular the transition rates $\tau_\pm$ for single individuals, defined in eq. (14). Using the simplified expression for $T_\pm$ from eq. (21) and recalling that $X_\pm = \frac{N}{2}(1 \pm z)$, one can write

$$\begin{cases} F(z) \approx (1-z)\tau_+(z) - (1+z)\tau_-(z) \\ D(z) \approx \frac{1}{N}\left[(1-z)\tau_+(z) + (1+z)\tau_-(z)\right]. \end{cases} \qquad (27)$$

One can see that the pre-factor $\frac{2}{N}$ in the expression for the drift coefficient $F(z)$, which appeared as a by-product of the first derivative $\partial_z$, is cancelled by the fact that the group-level transition rates $T_\pm$ increase linearly with $N$. Consequently, the drift coefficient – when written in terms of the individual transition probabilities $\tau_\pm$ – remains finite in the limit of infinitely many individuals.

The diffusion coefficient, on the other hand, acquires a pre-factor $\frac{4}{N^2}$ due to the second derivative $\partial_z^2$, which is only partially cancelled by the $N$-scaling of the transition rates $T_\pm$. Consequently, the diffusion coefficient, when written in terms of the individual transition rates $\tau_\pm$, vanishes as $1/N$ in the limit of infinitely many individuals. This is, in fact, not surprising. To see this, consider the example of $N$ completely non-interacting individuals, each of which independently changes direction with some fixed transition rate $\tau$: as a function of $N$, the number of individuals that change direction per time interval grows as $N$, but the net change in the number of aligned individuals – since most of the changes cancel out – scales only as $\sqrt{N}$. (It is equivalent to the expected traversed distance in an $N$-step random walk.) Moreover, changing the number of aligned individuals by 1 only changes the normalized parameter $z$ by $2/N$. Thus, the overall change to $z$ over a fixed time scales as $1/\sqrt{N}$, and the diffusion coefficient, which is related to the square of this parameter, shrinks as $1/N$ with the size of the population. For the purposes of our analysis, we are interested in the effects of interaction, which appear in addition to this simple scaling. They are captured by the rescaled diffusion coefficient

$$D'(z) \equiv ND(z), \qquad (28)$$

as used in the main text, section 4.1, eq. (5).

Finally, we note that, in the extrema $z = \pm 1$, the expressions for the drift and the (effective) diffusion coefficients reduce to

$$\begin{cases} F(z=-1) = \tau_+(z=-1) = D'(z=-1) \\ -F(z=+1) = \tau_-(z=+1) = D'(z=+1). \end{cases}$$

One can see this relation between drift and rescaled diffusion in the predictions of our model, shown in Fig. 2a. More importantly, the same can be seen in Fig. 2c, which depicts the predictions of a model that Dyson *et al.* fitted to experimental data [4].